\documentclass[journal,comsoc]{IEEEtran}

\usepackage{hyperref}
\usepackage[T1]{fontenc}
\usepackage[cmintegrals]{newtxmath}
\usepackage{cite}
\usepackage{bm}
\usepackage{changepage}
\usepackage{subfigure}
\usepackage{mathtools}
\usepackage[boxruled]{algorithm2e}
\usepackage{blindtext, graphicx}
\usepackage{relsize}
\usepackage{amssymb}
\usepackage{array}
\usepackage{longtable}
\usepackage[long,nocomma]{optidef}
\usepackage{framed}
\usepackage[table,xcdraw]{xcolor}
\usepackage[hyphenbreaks]{breakurl}

\newcommand{\name}{{\sf{BWRH}}}

\begin{document}

\title{On Minimizing the Completion Times of Long Flows over Inter-Datacenter WAN}

\author{
    \IEEEauthorblockN{Mohammad Noormohammadpour\textsuperscript{1}, Ajitesh Srivastava\textsuperscript{2}, Cauligi S. Raghavendra\textsuperscript{1,2}}\\
    \IEEEauthorblockA{\textsuperscript{1}Ming Hsieh Department of Electrical Engineering, University of Southern California (USC)\\ \textsuperscript{2}Department of Computer Science, University of Southern California (USC)}
}

\maketitle

\begin{abstract}
Long flows contribute huge volumes of traffic over inter-datacenter WAN. The Flow Completion Time (FCT) is a vital network performance metric that affects the running time of distributed applications and the users' quality of experience. Flow routing techniques based on propagation or queuing latency or instantaneous link utilization are insufficient for minimization of the long flows' FCT. We propose a routing approach that uses the remaining sizes and paths of all ongoing flows to minimize the worst-case completion time of incoming flows assuming no knowledge of future flow arrivals. Our approach can be formulated as an NP-Hard graph optimization problem. We propose \name, a heuristic to quickly generate an approximate solution. We evaluate \name~against several real WAN topologies and two different traffic patterns. We see that \name~provides solutions with an average optimality gap of less than $0.25\%$. Furthermore, we show that compared to other popular routing heuristics, \name~reduces the mean and tail FCT by up to $1.46\times$ and $1.53\times$, respectively.
\end{abstract}


\IEEEpeerreviewmaketitle

\section{Introduction}
Recently, dedicated inter-datacenter networks have been used by multiple organizations to connect dozens of their datacenters such as Google's B4 \cite{b4, b4andafter}, Facebook's Express Backbone \cite{facebook-express-backbone}, and Microsoft's Global WAN \cite{swan-backbone}. These networks are owned or leased by one organization, are relatively small with tens of edges, and can be managed in a logically centralized manner, for example, using frameworks such as SDN \cite{sdn}. This opens up new opportunities for global network-wide optimizations by combining the knowledge of traffic generated at the datacenters and control over traffic forwarding in the network. Therefore, in this paper, we revisit the well-known flow routing problem over inter-datacenter networks.

We focus on long flows which carry tremendous volumes of data over inter-datacenter networks \cite{b4, tempus}. They are usually generated as a result of replicating large objects such as search index files, virtual machine migration, and multimedia content. For instance, over Facebook's Express Backbone, about 80\% of flows for cache applications take at least 10 seconds to complete \cite{social_inside}. Besides, the volume of inter-datacenter traffic for replication of content and data, which generates many long flows, has been growing at a fast pace \cite{facebook-express-backbone}.

In general, flows are generated by different applications at unknown times to move data across the datacenters. Therefore, we assume that flows can arrive at the inter-datacenter network at any time and no knowledge of future flow arrivals. Every flow is specified with a source, a destination, an arrival time, and its total volume of data. The Flow Completion Time (FCT) of a flow is the time from its arrival until its completion.

We focus on minimizing the completion times of long flows which is a critical performance metric as it can significantly affect the overall application performance or considerably improve users' quality of experience. For example, in cloud applications such as Hadoop, moving data faster across datacenters can reduce the overall data processing time. As another example, moving popular multimedia content quickly to a regional datacenter via replication allows improved user experience for many local users. To attain this goal, routing and scheduling need to be considered together which can lead to a complex discrete optimization problem. In this paper, we only address the routing problem, that is, choosing a fixed path for an incoming flow given the network topology and the currently ongoing flows while making no assumptions on the traffic scheduling policy. We focus on single path routing which mitigates the undesirable effects of packet reordering.

A variety of metrics have been used for path selection over WAN including static metrics such as hop count and interface bandwidth, and dynamic metrics such as end-to-end latency which is a function of propagation and queuing latency, and current link bandwidth utilization \cite{tvlakshman, routing-metric}. While these metrics are effective for routing of short flows, they are insufficient for improving the completion times of long flows as we will demonstrate. Over inter-datacenter WAN where end-points are managed by the organization that also controls the routing \cite{b4, swan-backbone, facebook-express-backbone}, one can use routing techniques that differentiate long flows from short flows and use flow properties obtained from applications, including flow size information, to reduce the completion times of long flows.

\vspace{0.5em}
In this context, our paper makes the following contributions:

\begin{itemize}
    \setlength{\itemsep}{0.5em}
    \item Assuming no knowledge of future flow arrivals and no constraints on the network traffic scheduling policy, we propose to minimize the worst-case completion time of every incoming flow given the network topology, the currently ongoing flows' paths, and their remaining number of data units. For any given scheduling policy, we route the flows to minimize the worst-case flow completion time. We refer to this routing approach as the Best Worst-case Routing (BWR).
    
    \item BWR aims to select a minimum weight path for every incoming flow where a path's weight is defined as the total number of remaining data units of all the ongoing flows that have a common edge with the path. It can be shown that BWR is NP-Hard and finding an optimal solution requires examining all existing paths between the two ends of an incoming flow. We develop a heuristic, called \name, to quickly compute a route.
    
    \item We run extensive simulations to compare \name's performance with that of several other routing heuristics, including popular ones. We show that \name~improves the mean and tail completion times by up to $1.46\times$ and $1.53\times$, respectively, given various flow size distributions and scheduling policies. We also show that over multiple topologies and with different traffic patterns, \name's optimality gap is, on average, below $0.25\%$.
\end{itemize}

\section{System Model} \label{model}
We consider a general network topology with bidirectional links and equal capacity of one for all edges and assume an online scenario where flows arrive at unknown times in the future and are assigned a fixed path as they arrive. Each flow is divided into many equal size pieces (e.g., IP datagrams) which we refer to as data units. We also assume knowledge of the flow size (i.e., number of a flow's data units) for the new flow and the remaining flow size for all ongoing flows. Given an index $i$, every flow $F_i$ is defined with a source $s_i$, a destination $t_i$, an arrival time $\alpha_i$, and a total volume of data $\mathcal{V}_i$. In addition, each flow is associated with a path $P_i$, a finish time $\beta_i$ which is the time of delivery of its last data unit, and a completion time $c_i = \beta_i - \alpha_i$. Finally, at any moment, the total number of remaining data units of $F_i$ is $\mathcal{V}^r_i \le \mathcal{V}_i$.

Similar to multiple existing inter-datacenter networks \cite{b4, facebook-express-backbone, swan-backbone}, we assume the availability of logically centralized control over the network routing. A controller can maintain information on the currently ongoing long flows with their remaining data units and perform routing decisions for an incoming long flow upon arrival.

We employ a slotted timeline model where at each timeslot a single data unit can traverse any path in the network. In other words, we assume a zero propagation and queuing latency which we justify by focusing only on long flows. Given this model, if multiple flows have a shared edge, only one of them can transmit during a timeslot. We say two data units are competing if they belong to flows that share a common edge. Depending on the scheduling policy that is used, these data units may be sent in different orders but never at the same time. Also, if two flows with pending data units use non-overlapping paths, they can transmit their data units at the same time if no other flow with a common edge with either one of these flows is transmitting at the same timeslot.

\section{Best Worst-case Routing}
We aim to reduce long flows' completion times with no assumption on the scheduling policy for transmission of data units. To achieve this goal, we propose the following routing technique referred to as Best Worst-case Routing (BWR):

\vspace{0.3em}
\textbf{Problem 1.} \textit{Given a network topology $G(V,E)$ and the set of ongoing flows $\pmb{\mathrm{F}}=\{F_i, 1 \le i \le N\}$, we want to assign a path $P_{N+1}$ to the new flow $F_{N+1}$ so that the worst-case completion time of $F_{N+1}$, i.e., $\max(c_{N+1})$ is minimized.}

\vspace{0.3em}
Assuming no knowledge of future flows and given the described network model, since only a single data unit can get through any edge per timeslot, the worst-case completion time of a flow happens when the data units of all the flows that share at least one edge with the new flow's path go sequentially and before the last data unit of the new flow is transmitted. Therefore, Problem 1 can be reduced to the following graph optimization problem which aims to minimize the number of competing data units with $F_{N+1}$. 

\vspace{0.3em}
\textbf{Problem 2.} \textit{Given a network topology $G(V,E)$ where every edge $e \in E$ is associated with a set of flows $\pmb{\mathcal{\mathrm{F}}}_e$ (that is, $e \in P_{i}, \forall F_i \in \pmb{\mathcal{\mathrm{F}}}_e$), the set of ongoing flows $\pmb{\mathrm{F}}=\{F_i, 1 \le i \le N\}$, and an incoming flow $F_{N+1}$, we want to find a minimum weight path $P_{N+1}$ where the weight of any path $P$ from $s_{N+1}$ to $t_{N+1}$ is computed as follows:}

\vspace{-0.5em}
\begin{equation}
    W_{P} = \sum_{\{1 \le i \le N ~\vert~ F_i~\in~\{\cup_{e \in P}~\pmb{\mathcal{\mathrm{F}}}_e\}\}} \mathcal{V}^r_i \label{eq1}
\end{equation}

\vspace{0.3em}
\textbf{Proposition 1.} \textit{Assuming no knowledge of future flow arrivals, $P_{N+1}$ selected by solving Problem 2 minimizes the worst-case completion time of $F_{N+1}$ regardless of the scheduling policy used for transmission of data units.}

\vspace{0.3em}
\textbf{\textit{Proof.}} $P_{N+1}$ is chosen to minimize the maximum number of data units ahead of $F_{N+1}$ given the knowledge of ongoing flows' remaining data units which minimizes the worst-case $\beta_{N+1}$, that is, the maximum number of timeslots the last data unit of $F_{N+1}$ has to wait before it can be sent. Since $\alpha_{N+1}$ is fixed, this minimizes $\max(c_{N+1})$.

\begin{figure}[t]
    \centering
    \includegraphics[width=\columnwidth]{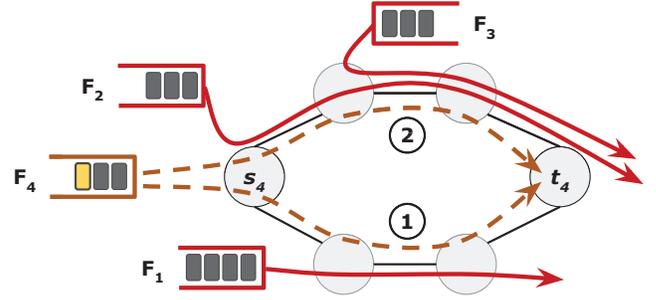}
    \caption{Example of routing a new flow $F_4$} \label{fig:example}
\end{figure}

\vspace{0.3em}
\textbf{Example:} Consider the scenario shown in Figure \ref{fig:example}. A new flow $F_4$ with 3 data units has arrived and has two options of sharing an edge with $F_1$ that has 4 remaining data units (path 1) or sharing edges with $\{F_2, F_3\}$ which have a total of 6 remaining data units (path 2). Our approach tries to minimize the worst-case completion time of $F_4$ given ongoing flows. If path 1 is chosen, the worst case completion time of $F_4$ will be 7 while with path 2 it will be 9 and therefore, the logically centralized network controller will select path 1 for $F_4$. The worst-case completion times are not affected by the scheduling policy and are independent of it. Also, the fact that $F_2$ has three common edges with path 2 and $F_3$ has two common edges with path 2 does not affect the worst-case completion time of $F_4$ on path 2.

\section{BWR Heuristic (\name)}
The path weight assignment used in Problem 2 is not edge-decomposable. Finding a minimum weight path for $F_{N+1}$ is NP-Hard and requires examining all paths from $s_{N+1}$ to $t_{N+1}$.\footnote{NP-Hardness proof dropped for brevity. It can be shown that the Set Cover problem can be reduced to Problem 2. In particular, we are looking for a subset of flows $\gamma \subseteq \pmb{\mathrm{F}}$ with minimum total remaining data units where there exists a path from $s_{N+1}$ to $t_{N+1}$ removing all edges that have a flow in $\pmb{\mathrm{F}}-\gamma$.} 

We propose a fast heuristic here, called \name, that finds an approximate solution to Problem 2. Algorithm \ref{bwrh} shows our proposed approach to finding a path $P_{N+1}$ for $F_{N+1}$. At every iteration, the algorithm finds the minimum weight path from $s_{N+1}$ to $t_{N+1}$ with at most $K$ hops by computing the weight of every such path according to Eq. \ref{eq1}. The algorithm starts by searching all the minimum hop paths from $s_{N+1}$ to $t_{N+1}$ and finding the weight of the minimum weight path among such paths. It then increases the number of maximum hops allowed (i.e., $K$) by one, extending the search space to more paths. This process continues until the weight of the minimum weight path with at most $K$ hops is the same as $K-1$, i.e., there is no gain while increasing the number of hops.

The termination condition used in \name~may prevent us from searching long paths. Therefore, if the optimal path is considerably longer than the minimum hop path, it is possible that the algorithm terminates before it reaches the optimal path. Let us call the optimal path $P_o$ and the path selected with our heuristic $P_h$. The optimality gap, defined as $\frac{W_{P_h} - W_{P_{o}}}{W_{P_{o}}}$, is highly dependant on the number of remaining data units of ongoing flows. We find that the worst-case optimality gap can be generally unbounded. However, it is highly unlikely, in general, for the optimal path to be long as having more edges increases the likelihood of sharing edges with more ongoing flows which increases the weight of the path. We will later confirm this intuition through empirical evaluations and show that \name~provides solutions with an average optimality gap of less than a quarter of percent.

\SetAlgoVlined
\begin{algorithm}[t]
\caption{\name} \label{bwrh}
{\small
\SetKw{KwBy}{by}
\SetKwProg{FindPath}{FindPath}{}{}

\vspace{0.4em}
\KwIn{$F_{N+1}$, $G(V,E)$, $P_i,\mathcal{V}^r_i,1 \le i \le N$}

\vspace{0.4em}
\KwOut{$P_{N+1}$}

\hrulefill

\vspace{0.4em}
$K \gets$ $\#$hops on the minimum hop path from $s_{N+1}$ to $t_{N+1}$\;

\vspace{0.4em}
$W_{min}^K \gets$ Weight of the minimum weight path from $s_{N+1}$ to $t_{N+1}$ with at most $K$ hops by examining all such paths\;

\vspace{0.4em}
\Repeat{$W_{min}^{K} \ge W_{min}^{K-1}$}{
    \vspace{0.4em}
    $K \gets K+1$\;
    
    \vspace{0.4em}
    Compute $W_{min}^K$\;
    
    \vspace{0.4em}
}

\vspace{0.4em}
$P_{N+1} \gets$ The minimum weight path from $s_{N+1}$ to $t_{N+1}$ with at most $K-1$ hops (if multiple minimum weight paths exist, choose the one with minimum hops)\;
}
\end{algorithm}

\section{Application to Real Network Scenarios}
We discuss how \name~can be used to find a path for an incoming flow on a real network assuming a uniform link capacity. We can use the same topology as the actual topology as input to \name. Since we focus on long flows for which the transmission time is significantly larger than both propagation and queuing latency along existing paths, it is reasonable to ignore their effect in routing (hence the assumption that these values are zero in \S \ref{model}). Next, assuming that all data units are of the same size, we can use the total number of remaining bytes per ongoing flow in place of the number of remaining data units as it does not affect the selected path. In practice, some data units may be smaller than the underlying network's MTU, which for the long flows with many data units, has minimal effect on the selected path. Once \name~selects a path, the network's forwarding state is updated accordingly to route the new flow's traffic, for example, using SDN \cite{b4, tempus}.

In general, network traffic is a mix of short and long flows. Since our proposal targets the long flows, routing of short flows will not be affected and could be done considering the propagation and queuing latency. Incoming long flows can be routed according to the knowledge of current long flows while ignoring the effect of short flows. Improving the completion times of long flows in a network with a mix of short and long flows is part of the future work.

\section{Evaluations}
We considered two flow size distributions of light-tailed (Exponential) and heavy-tailed (Pareto) and considered Poisson flow arrivals with the rate of $\lambda$. We also assumed an average flow size of $\mu$ data units with a maximum of 500 data units along with a minimum size of 2 data units for the heavy-tailed distribution. We considered the scheduling policies of First Come First Serve (FCFS), Shortest Remaining Processing Time (SRPT) and Fair Sharing based on max-min fairness \cite{max-min-fairness}.

\vspace{0.3em}
\textbf{Topologies:} GScale \cite{b4} with 12 nodes and 19 edges, AGIS \cite{agis} with 25 nodes and 30 edges, ANS \cite{ans} with 18 nodes and 25 edges, and Cogent \cite{cogent} with 197 nodes and 243 edges. We assumed bidirectional edges with a uniform capacity of 1 data unit per time unit for all of these topologies. 

\vspace{0.3em}
\textbf{Schemes:} We considered three schemes besides \name. The \textit{Shortest Path (Min-Hop)} approach simply selects a fixed shortest hop path from the source to destination per flow. The \textit{Min-Max Utilization} approach selects a path that has the minimum value of maximum utilization across all paths going from the source to the destination. This approach has been extensively used in the traffic engineering literature \cite{tvlakshman, tempus}. The \textit{Shortest Path (Random-Uniform)} selects a path randomly with equal probability across all existing paths which are at most one hop longer than the shortest hop path.

\begin{figure}[b]
    \centering
    \includegraphics[width=\columnwidth]{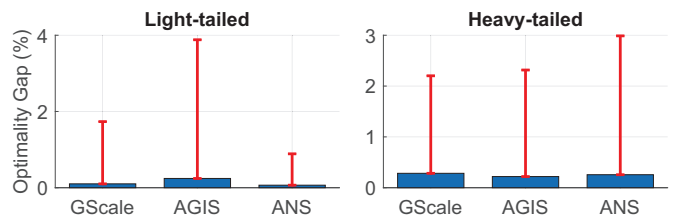}
    \caption{\name's optimality gap for $\lambda = 10$ and $\mu = 50$ computed for 1000 flow arrivals.} \label{fig:exp0}
\end{figure}

\begin{figure*}[t]
    \centering
    \includegraphics[width=0.95\textwidth]{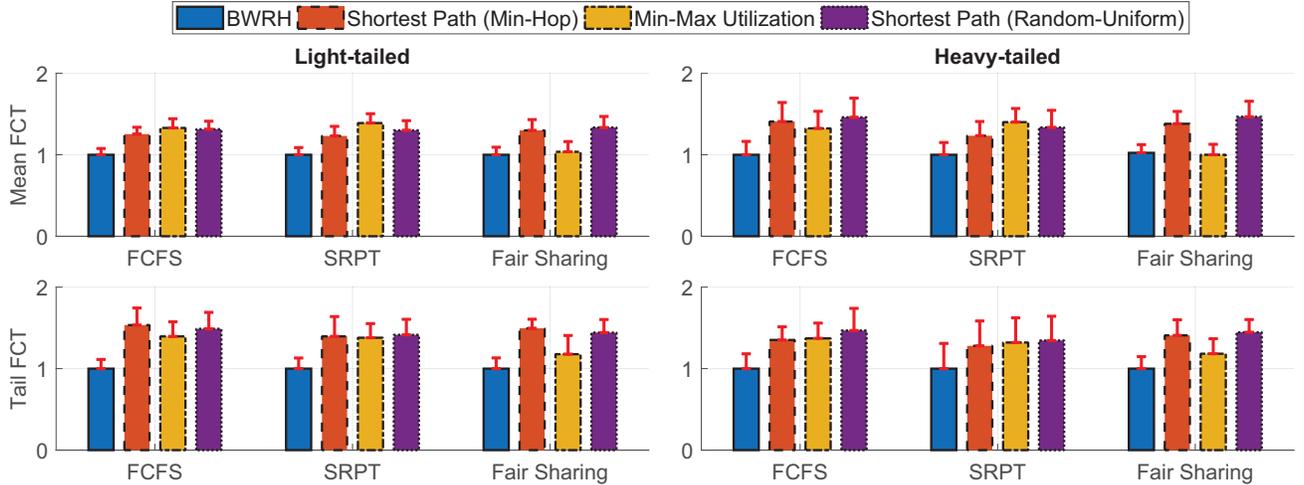}
    \caption{Online routing techniques by flow scheduling policy assuming $\lambda=1$, $\mu=50$ and Cogent \cite{cogent} topology over 500 time units.} \label{fig:exp2}
\end{figure*}

\vspace{0.3em}
\textbf{\name's Optimality Gap:} In Figure \ref{fig:exp0} we compute the optimality gap of solutions found by \name~over three different topologies and under two traffic patterns. The optimal solution was computed by taking into account all existing paths and finding the minimum weight path on topologies of GScale, AGIS, and ANS. We also implemented a custom branch and bound approach which would require less computation time with a small number of ongoing flows (i.e., $< 20$ in our setting) and an intractable amount of time for a large number of ongoing flows (i.e., $> 30$ in our setting). According to the results, the average gap is less than $0.25\%$ over all experiments. We could not perform this experiment on larger topologies as computing the optimal solution would take an intractable amount of time.

\vspace{0.3em}
\textbf{Effect of Scheduling Policies:} In Figure \ref{fig:exp2} we fixed the flow arrival rate to 1 and mean flow size to 50 and tried various scheduling policies under the Cogent topology which is much larger than GScale, ANS, and AGIS. All simulations were repeated 10 times and the standard deviation for each instance has been reported. The minimum value normalizes each group of bars. We see that \name~is consistently better than other schemes regardless of the scheduling policy used. We can also see that compared to each other, the performance of other schemes varies considerably with the scheduling policy applied. To quantify, \name~provides between $1.18\times$ to $1.53\times$ better tail completion times than the other schemes across all scenarios on average. We also observe up to $1.46\times$ better mean completion times compared to other schemes across all scheduling policies on average.

\vspace{0.3em}
\textbf{Running Time:} We implemented Algorithm \ref{bwrh} in Java using the JGraphT library. To exhaustively find all paths with at most $K$ hops, we used the class \texttt{AllDirectedPaths} in JGraphT. We performed simulations while varying $\lambda$ from 1 to 10 and $\mu$ from 5 to 50 over 1000 flow arrivals per experiment which covers both lightly and heavily loaded regimes. We also experimented with all the four topologies pointed to earlier, both traffic patterns of light-tailed and heavy-tailed, and all three scheduling policies of FCFS, SRPT, and Fair Sharing. The maximum running time of Algorithm \ref{bwrh} was $222.24$ milliseconds, and the average of maximum running time across all experiments was $27$ milliseconds. This latency can be considered negligible given the time needed to complete long flows once they are routed.

\section{Conclusions and Future Work}
We proposed a new technique for routing based on flow size information to reduce flow completion times. Accordingly, the online routing problem turns into finding a minimum weight path on the topology from the source to the destination where the weight is computed by summing up the number of remaining data units of all the flows that have a common edge with the path. Since this is a hard problem, we proposed a fast heuristic with small average optimality gap. We then discussed how information from a real network scenario could be used as input to the proposed network model to find a path on the actual network for an incoming flow. In the future, we would like to study networks with non-uniform link capacity, multipath routing, and the effect of inaccurate flow size information on routing performance.

{\bibliography{citations.bib}
\bibliographystyle{IEEEtran}}
\end{document}